\documentclass[a4,aps,showpacs,amsmath,floatfix,twocolumn]{revtex4}
\usepackage{amsmath, bm}
\usepackage{graphicx}
\usepackage{dcolumn}
\usepackage{dirtytalk}
\usepackage{color}
\usepackage{amsmath}
\usepackage{here}
\usepackage[utf8x]{inputenc}
\usepackage{bm}
\usepackage[mathscr]{euscript}
\usepackage{amssymb}
\usepackage{amsmath}

\usepackage{hyperref}
\hypersetup{breaklinks=true,
            colorlinks=false,
            pdfborder={0 0 0}}
\usepackage{bm}
\usepackage{braket}
\usepackage{MnSymbol}
\usepackage{graphicx}
\graphicspath{{figures/}}
\usepackage{dsfont}

\usepackage{titlesec}

\begin{document}
\title{The Berry Phase Rectification Tensor and The Solar
Rectification Vector}

\author{O. Matsyshyn$^{1}$, Urmimala Dey$^{1,3}$, I. Sodemann$^{1}$, Y. Sun$^{2}$}

\affiliation{$^{1}$Max Planck Institute for the Physics of Complex Systems, Dresden 01187 , Germany
}
\affiliation{$^{2}$Max Planck Institute for Chemical Physics of Solids, Dresden 01187 , Germany
}
\affiliation{$^{3}$Centre for Theoretical Studies, Indian Institute of Technology Kharagpur, Kharagpur 721302, India
}
\begin{abstract}
We introduce an operational definition of the Berry Phase Rectification Tensor as the second order change of polarization of a material in response to an ideal short pulse of electric field. Under time reversal symmetry this tensor depends exclusively on the Berry phases of the Bloch bands and not their energy dispersions, making it an intrinsic property to each material which contains contributions from both the inter-band shift currents and the intra-band Berry Curvature Dipole. We also introduce the Solar Rectification Vector as a technologically relevant figure of merit for bulk photo-current generation under ideal black-body radiation in analogy with the classic solar cell model of Shockley and Queisser. We perform first principle calculations of the Berry Phase Rectification Tensor and the Solar Rectification Vector for the Weyl semimetal TaAs and the insulator LiAsSe$_2$ which features large shift currents close to the peak of solar radiation intensity.
\end{abstract}


\maketitle

\textit{\color{blue} Introduction}. Several remarkable relations between the quantum geometry and topology of electronic states in crystalline solids and their non-linear optical and transport properties have been unraveled over the last few decades. One of the second order effects intimately intertwined with the geometry of electronic wave-functions is the shift current \cite{PhysRevB.19.1548,PhysRevB.23.5590,Sturman1992,PhysRevB.52.14636,PhysRevB.61.5337,BrehmYoung,Morimotoe1501524,NaMo,PhysRevB.99.045121,PhysRevLett.123.246602}, which is a rectified current arising in response to AC electric fields that drive optical transitions in crystals without inversion symmetry. The shift current arises from band-off-diagonal components of the velocity operator, in contrast to the injection current, which arises from the band-diagonal difference of group velocities between empty and occupied states \cite{Sturman1992,PhysRevB.23.5590,PhysRevB.61.5337,Morimotoe1501524,NaMo,PhysRevB.99.045121,PhysRevLett.123.246602,BrehmYoung}. In materials with time reversal symmetry, the shift current differs from the injection current in that only the shift current can lead to rectified currents in response to linearly polarized electric fields, making it attractive for photovoltaic applications \cite{BrehmYoung,PhysRevLett.109.116601,PhysRevLett.119.067402,Cook2017,PhysRevLett.121.267401} since it can lead to a DC current for electric fields with no net degree of circular polarization such as those arriving from the sun.

\begin{figure}[t]
    \centering
    \includegraphics[width=0.48\textwidth]{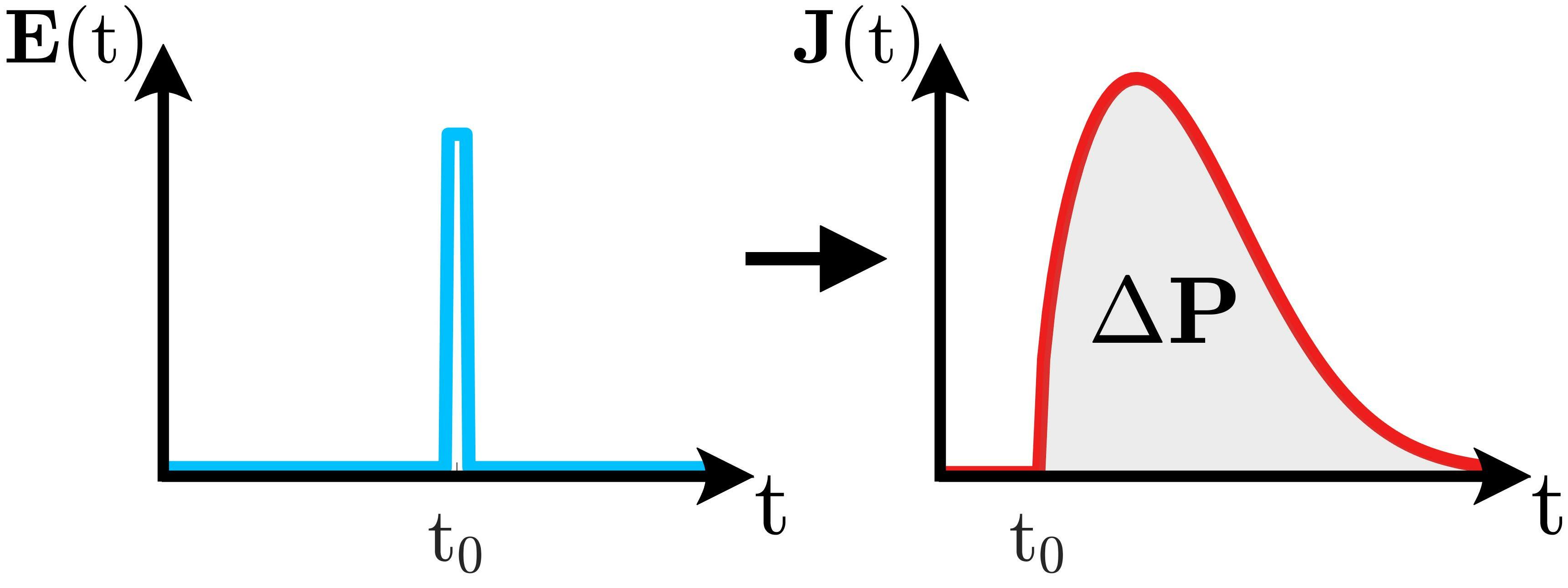}
    \caption{The Berry Phase Rectification Tensor measures the second order change of polarization (which equals the area in the right panel) in response to delta function pulse of electric field (left panel).}
    \label{fig:FIG1}
\end{figure}

Another second order effect with an intimate connection to the geometry of electronic wave functions is the non-linear Hall effect driven by the Berry Curvature Dipole (BCD) \cite{2009arXiv0904.1917D,PhysRevLett.123.246602,PhysRevLett.105.026805,PhysRevLett.115.216806}, that has been recently observed experimentally \cite{2018arXiv180908744K,Ma2019,2019arXiv190202699S,PhysRevLett.123.036806,2019arXiv191007509H,2020arXiv200605615H}. Like the shift current, the BCD non-linear Hall effect can also give rise to a rectified
current in response to a linearly polarized electric field, but with the crucial difference that the BCD effect occurs only in metals at low frequencies in a Drude-like fashion \cite{PhysRevLett.123.246602}. These two distinct effects, however, appear to have a connection, as suggested by the quantum-rectification sum rule (QRSR) derived in Ref.\cite{PhysRevLett.123.246602}. The QRSR states that the frequency integral of the total rectification conductivity for any time-reversal-invariant material is completely independent of energy dispersions and entirely controlled by the quantum geometric Berry connections of the bands. The contributions to the QRSR arise from the BCD effect, which accounts for all the intra-band weight in a Drude-like peak, and the shift current, which accounts for all the inter-band contributions. In contrast, injection currents do not contribute to the QRSR under time reversal invariant conditions.

\begin{figure}[t]
    \centering
    \includegraphics[width=0.48\textwidth]{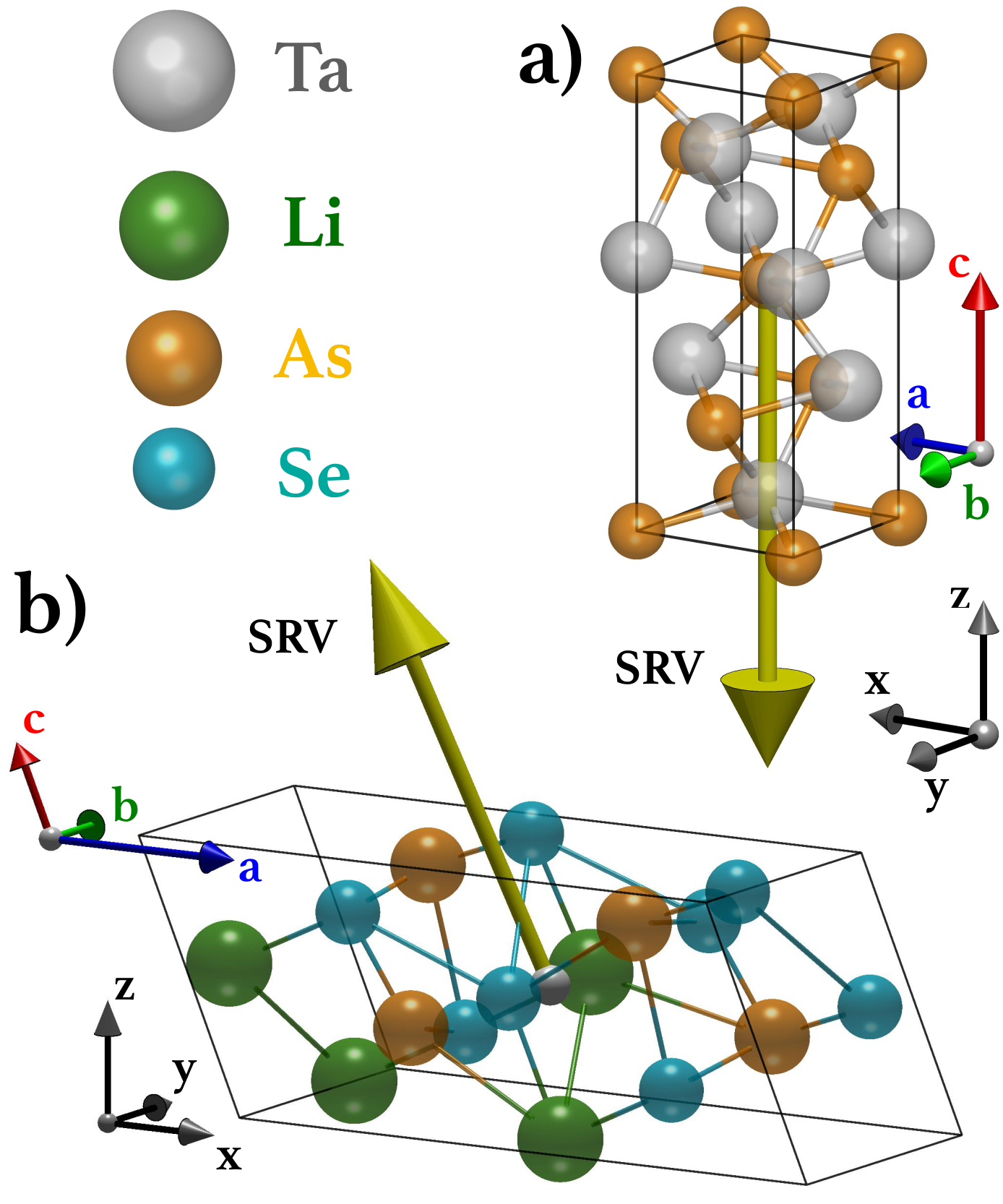}
    \caption{Crystal structures and Solar Rectification Vectors (SRV) for (a) TaAs and (b) LiAsSe$_2$. }
    \label{fig:sori}
\end{figure}

In this work, we will show that the QRSR defines a three-index Berry Phase Rectification Tensor (BPRT), which depends only on the Berry phase geometry of a time-reversal-invariant material. In d-dimensions the BPRT has units of $\mathrm{(length)}^{d-3}$ , and therefore,
it is dimensionless for 3D materials. We will introduce an operational definition of the BPRT by considering the change of polarization in response to ideal short pulses of electric fields (see Fig.\ref{fig:FIG1}) and compute it from first principle calculations for the Weyl semi-metal TaAs \cite{PhysRevX.5.011029,Lv2015,Huang2015,Xu613} and demonstrate that, remarkably, the net contribution from the BCD \cite{PhysRevB.97.041101} is comparable to the inter-band contribution from shift currents. We will also compute the BPRT for the insulator LiAsSe$_2$, which is a promising photo-voltaic material \cite{BrehmYoung}. Additionally we will introduce the notion of the Solar Rectification Vector (SRV), whose magnitude serves as a technologically useful figure of merit for bulk photo-current generation, and is defined from the average DC current produced under irradiation from ideal black-body model of sun-light, analogous to that employed by Shockley and Quessier (SQ) in their seminal study of p-n junction solar cells \cite{ShockleyQueisser}. We will demonstrate that LiAsSe$_2$ could generate photo-currents as large as 8 $\mathrm{mA/cm^2}$, which is about 8$\%$ of the ideal maximum from the SQ model  \cite{ShockleyQueisser}. The direction of the  SRV for these materials is shown as a yellow vector in Fig.[\ref{fig:sori}].

\textit{\color{blue}The Quantum Rectification Tensor}. Consider an electronic system which is in equilibrium for $t<0$, and at $t=0$ is subjected to a quench of
the vector potential, leading to a delta-function pulse in the electric field of the form:
\begin{equation}
    \mathbf{A}(t) = \mathbf{A}_0\Theta(t), \qquad \mathbf{E}(t)=\mathbf{E}_0\delta(t),
\end{equation}
where $\Theta(t)$ is the heavy-side function. After the pulse, the net change of electric polarization can be obtained from:
\begin{equation}
    \Delta\mathbf{P} = \int_0^\infty \mathbf{j}(t) dt.
\end{equation}
This protocol is illustrated in Fig.[\ref{fig:FIG1}]. To second order in electric fields the current is related to the linear and  second order conductivity as:
\begin{multline}\label{currenttot}
    j^\gamma(t) = \int_{-\infty}^{\infty}dt_1\sigma_{(1)}^{\gamma\alpha}(t,t_1)E^\alpha(t_1)+\\+\int_{-\infty}^{\infty}dt_1\int_{-\infty}^{\infty}dt_2\sigma_{(2)}^{\gamma\alpha\beta}(t,t_1,t_2)E^\beta(t_1)E^\alpha(t_2).
\end{multline}
where Greek letters denote space indices and here and throughout the paper sum over repeated indices is understood unless otherwise stated.
Therefore, the expansion of the polarization in powers of $A_0$ is:
\begin{equation}\label{dp}
    \Delta P^\gamma = -\sigma^{\gamma\alpha}A_0^\alpha+\frac{e^3}{\hbar^2}R^{\gamma\alpha\beta}A_0^{\beta}A_0^{\alpha} + \ldots,
\end{equation}
where $\sigma^{\gamma\alpha}$ is the zero frequency linear conductivity, and $R^{\gamma\alpha\beta}$ defines the Berry Phase Rectification Tensor (BPRT) of the system, which is given by:
\begin{equation}
    \frac{e^3}{\hbar^2}R^{\gamma\alpha\beta} = \int_{-\infty}^{\infty}dt \sigma_{(2)}^{\gamma\alpha\beta}(t,0,0)=\int_{-\infty}^{\infty}\frac{d\omega}{2\pi} \sigma_{(2)}^{\gamma\alpha\beta}(-\omega,\omega).
\end{equation}

Eq.(\ref{dp}) expresses the notion that the BRPT measures the net displacement of the average position of an electronic system to second order in the strength of delta function pulse in electric field. As demonstrated in \cite{PhysRevLett.123.246602} for the ground state of any non-interacting system with a time-reversal-invariant band structure, $R^{\gamma\alpha\beta}$ only contains information about the Berry connections of the bands. Specifically, in the clean limit of a time reversal invariant band structure, it is given by:
\begin{multline}\label{eq6}
    R^{\gamma\alpha\beta} = \frac14\Big\{\left\langle\partial^\alpha\Omega^{\beta\gamma}\right\rangle+
    \left\langle\left[A^{\alpha},i\partial^{\beta}A^{\gamma}\right]\right\rangle+\\+\left\langle\left[A^\alpha,\left[A^\beta,\bar{A}^\gamma\right]\right]\right\rangle+\left(\alpha\leftrightarrow \beta\right)\Big\},
\end{multline}
where $\partial^\alpha \equiv \partial/\partial k^\alpha$, ${\Omega}_{n}^{\alpha\beta}= \partial^\alpha {A}_{nn}^{\beta}-\partial^\beta {A}_{nn}^{\alpha}$, ${\bar{A}}^{\alpha}_{nm} ={{A}}^{\alpha}_{nm}(1-\delta_{nm})$ is the off-diagonal non-Abelian Berry connection and average is defined as:

\begin{equation}
    \langle \cdots \rangle = \sum_n \int \frac{d^d\mathbf{k}}{(2\pi)^d}f_n\langle n \mid(\cdots)\mid n\rangle,
\end{equation}
where $f_n$ is the Fermi-Dirac occupation function of the $n$-th Bloch band. The first term in Eq. (\ref{eq6}) arises from the intra-band BCD while the two other terms arise from the inter-band shift current. Therefore the BPRT connects these two effects and allows for a concrete measure of the shift current effect, as a collective displacement of average position of the electron system in response to an electric field pulse. Such an interpretation of the BPRT suggests that time domain spectroscopic techniques ~\cite{Kohli11,2020arXiv200511550A} are promising tools to measure the BPRT in materials. Notice that by construction the BPRT is symmetric in its last two indices, and the BCD does not contribute to fully diagonal components of the form $R^{\alpha\alpha\alpha}$. For a related, but more restricted, sum rule derived see Ref.\cite{PhysRevB.98.165113}. 

\textit{\color{blue}The Solar Rectification Vector}. In this section we introduce a figure of merit for bulk photovoltaic materials that captures their
amount of DC current generated from sun-light under ideal conditions, and therefore it is a more relevant quantity for technological applications. Consider an statistical
ensemble average of the electric field in Eq.(\ref{currenttot}), given by:
\begin{figure}[t]
    \centering
    \includegraphics[width=0.48\textwidth]{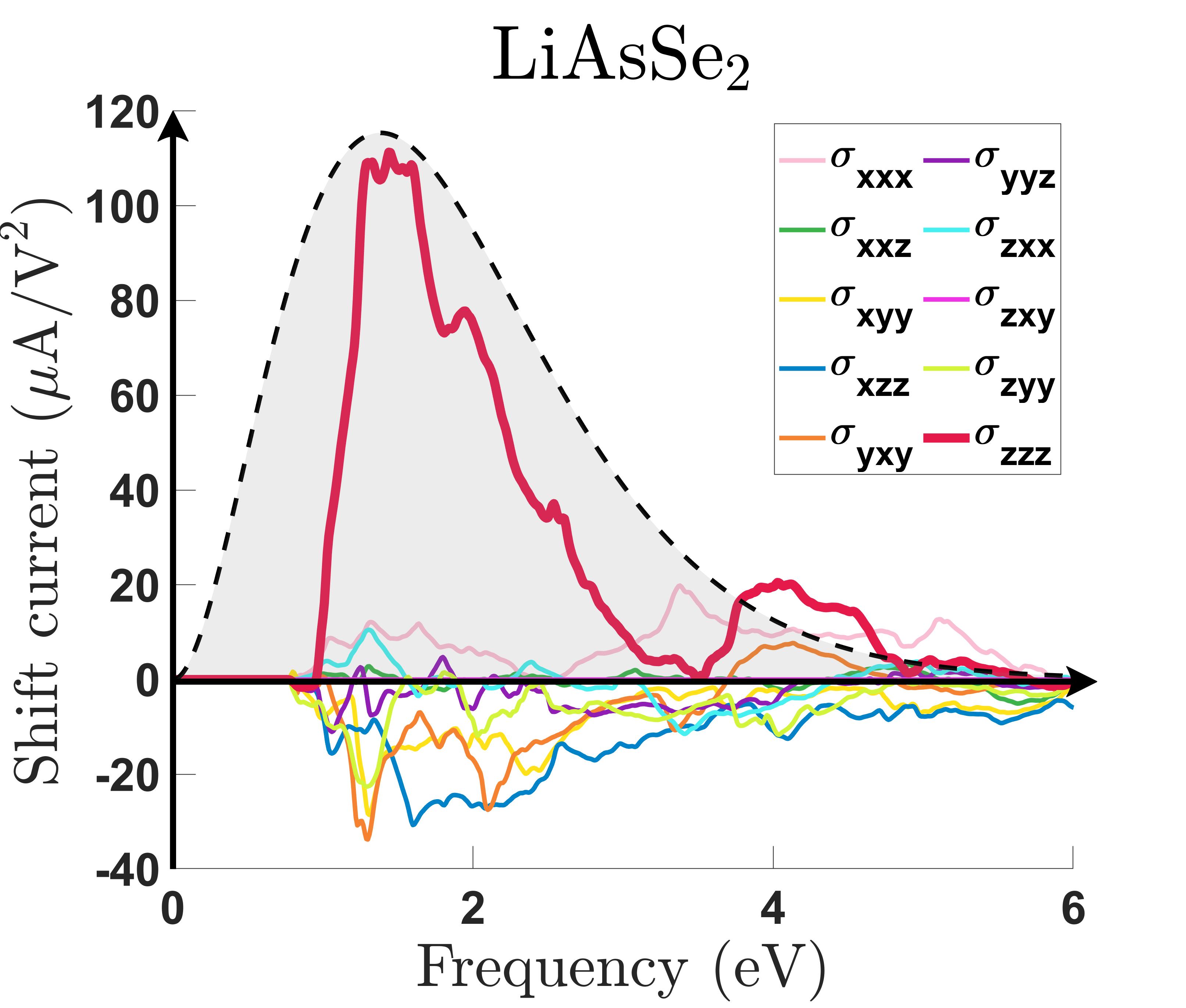}
    \caption{The non-zero conductivity components for LiAsSe$_2$ as a function of frequency. The dashed line is the solar Planck intensity distribution (in arbitrary units) showing its good overlap with the $\sigma^{zzz}_{(2)}$ rectification conductivity of LiAsSe2. }
    \label{fig:fig3}
\end{figure}
\begin{equation}
    \overline{j^\gamma(t)} =\iint_{-\infty}^{\infty}dt_1dt_2\sigma_{(2)}^{\gamma\alpha\beta}(t,t_1,t_2)\overline{E^\beta(t_1)E^\alpha(t_2)}.
\end{equation}
Assuming time translational invariance of the Hamiltonian and the ensemble, we have: 
\begin{equation}\label{autocoorel}
    \overline{E^\beta(t_1)E^\alpha(t_2)}=I^{\beta\alpha}(t_1-t_2).
\end{equation}
Then the average current will be time independent and given by:
\begin{equation}
    \overline{j^\gamma} =\int_{-\infty}^{\infty}d\omega\sigma_{(2)}^{\gamma\alpha\beta}(-\omega,\omega)I^{\beta\alpha}(\omega).
\end{equation}
Where $I^{\beta\alpha}(\omega)$ is the Fourier transform of the electric field auto-correlation function. For the ensemble of black-body radiation one obtains that (see S.I. \ref{BBrad}):
\begin{equation}\label{corelfun}
    I^{\beta\alpha}(\omega)=\frac{\hbar\delta^{\beta\alpha}}{(2\pi)^2\epsilon_0(\beta_S\hbar c)^3}\left(\frac{R_{\mathbf{S}}}{R_\mathbf{E}}\right)^2\mathcal{I}(\hbar\beta_S|\omega|),
\end{equation}
where $\mathcal{I}(x)= x^3/(e^x-1)$ is the Planck spectral distribution function and $\beta_S = 1/(k_BT_S)$, is the inverse temperature of the sun $T_S \approx 5778$K, and $R_\mathbf{S}$ is the radius of the sun, and $R_\mathbf{E}$ is the sun to earth distance. Therefore the electric current density per photon flux (number of photons per unit
area per unit time) $F_S$ ([$F_S$] = $m^{ -2} s^{-1}$ ), can be written as:
\begin{equation}\label{SQsm}
    j^\gamma = e F_s\eta^\gamma.
\end{equation}
Here ${\eta}^\gamma$ defines the Solar Rectification Vector (SRV), which depends only on the temperature of the sun, and
the intrinsic rectification properties of the material and is explicitly given by:
\begin{equation}\label{etagamma}
    \eta^\gamma = \frac{\hbar}{2\zeta(3)ec\epsilon_0}\int_{-\infty}^{\infty}d\omega~ \sigma_{(2)}^{\gamma\alpha\alpha}(-\omega,\omega)\mathcal{I}(\hbar\beta_S|\omega|),
\end{equation}
here $\zeta$ is the Riemann zeta-function. Notice that $\eta^\gamma$ is a dimensionless vector for 3D materials and its magnitude, $\eta = \sqrt{\sum_i\eta_i^2}$, counts the number of electrons contributing to the rectified current per incoming photon, and serves as a dimensionless figure of merit for the ideal bulk photo-current generated by a material under short-circuit configurations. Because the Planck distribution has vanishing intensity at small frequency, in the clean limit, the BCD Drude-like peak does not contribute to the rectified current and all the contributions arise from shift currents in time-reversal invariant materials, while injection current contributions vanish (see S.I. \ref{SumRuleComp}). 

It is instructive to contrast this figure of merit with a comparable figure of merit for current generation for the ideal model of a p-n junction solar-cell proposed by Shockley-Quessier (SQ) \cite{ShockleyQueisser}. In the most ideal version of the SQ model one assumes that every photon with energy above the band-gap is absorbed and generates an electron and a hole that are successfully collected at opposite electrodes contributing to the current generation of the material. Therefore in this model the total current density is:

\begin{equation}\label{jsq}
    j_{\mathrm{SQ}} = e F_s r.
\end{equation}
Here $r$ is the fraction of the total number of incident photons that have energy above the band-gap and thus can be absorbed by the material (see S.I.~\ref{BBrad}). SQ were primarily interested in optimizing the total power delivered by the solar cell and not the total current. It is however clear that the maximum value of $r$ is simply $r =1$ which is achieved in the limit of vanishing gap. As we will see LiAsSe$_2$ has a value of $\eta =
0.08$, and thus can produce a photocurrent that is $8\%$ of this ideal limit.

Our discussion of photo-current generation so far has assumed that light propagates inside the material as it would do in vacuum. However, a more realistic calculation of the photo-current would  include the frequency dependence of light absorbtion and propagation in the material, which is often accounted for by the Glass coefficient~\cite{Glass,PhysRevLett.109.116601}. Assuming perpendicular incident light along the z-axis, that the material has crystal symmetry so that its linear conductivity is a diagonal tensor, and that the material is thick enough to absorb all light above its optical gap, one obtains a total current given by (see S.I. \ref{GlassSupp}):
\begin{equation}\label{igamma}
   \overline{I^\gamma} =W\int_{-\infty}^{\infty}d\omega |T^\beta(\omega)|^2\frac{\sigma_{(2)}^{\gamma\beta\beta}(-\omega,\omega)}{\alpha^\beta(\omega)}I^{\beta\beta}(\omega),
\end{equation}
where $W$ is the sample width, $T^\beta(\omega) =2n_1/(n_1+n^\beta_2(\omega))$, $\alpha^\beta(\omega)=2\mathrm{Im}[n^\beta_2(\omega)]\omega/c$, $c$ is the speed of light, $n_1$ is an index of vacuum refraction, $n^\beta_2$ is direction dependent index of the material of interest refraction, the function $I^{\beta\beta}(\omega)$ is given in Eq.(\ref{corelfun}) and the index $\beta$ here is restricted to take values $\beta=x,y$. The formula above generalizes that from Refs.~\cite{Glass,PhysRevLett.109.116601} by adding the frequency dependence of the reflection coefficient at the material surface, the spectral distribution of incident light as well as the directional dependence of the conductivity of the material. We expect it to provide accurate estimates in bulk-photovoltaic solar cell devices.

\textit{\color{blue}Bulk rectification of TaAs and LiAsSe$_2$}. In this section we combine our general considerations with first principle calculations of the non-linear opto-electronic properties of the 3D Weyl semi-metal TaAs and the insulator LiAsSe$_2$. Because it is semi-metallic, TaAs has BCD and shift current contributions whereas LiAsSe$_2$ only has shift current contributions. Our primary interest in TaAs is to investigate the interplay between the intraband BCD and interband shift-currents to its BPRT, although we also compute its QRV, while in the case of LiAsSe$_2$ our primary interest is the estimation of its SRV due to its promise as a photovoltaic material, but we also compute its BPRT.

The band structures of LiAsSe$_2$ and TaAs were calculated by the full-potential local-orbital (FPLO) \cite{PhysRevLett.77.3865,PhysRevB.59.1743} minimum-basis code with the experimental crystal structure. Details of the calculations of band structure, Berry phases and non-linear rectification conductivity are presented in S.I. \ref{DFTdetails}. The frequency dependent non-linear conductivity of LiAsSe$_2$ is shown in Fig.[\ref{fig:fig3}] (for the one of TaAs see S.I. \ref{DFTdetails}). Our results are in agreement with previous reports  \cite{BrehmYoung,PhysRevB.97.241118}. In particular, the $\sigma^{zxx}$ and $\sigma^{zzz}$ components for TaAs can exceed 300$\mu \mathrm{A/V}^2$ at infrared frequencies, while the $\sigma^{zzz}$ component of LiAsSe$_2$ can exceed 100$\mu \mathrm{A/V}^2$ in a wide range frequency from $\sim$1.25 to $\sim$1.65 eV, which as we will see, leads to a sizable SRV.
The space symmetry I4$_1$md of TaAs leads to the following non-zero components of the BPRT:
\begin{equation}
    \begin{array}{c}
        R^{xxz} = R^{yyz}=R^{xzx}=R^{yzy}=0.25,   \vspace{1mm}\\R^{zxx} = R^{zyy}=0.09,\qquad R^{zzz}=-0.07.
    \end{array}
\end{equation}
The values above contain the contribution from BCD and shift currents. The BCD part of the  contribution to BPRT is:
\begin{equation}
    \begin{array}{c}
          R^{xxz}_{BCD} = R^{yyz}_{BCD}=R^{xzx}_{BCD}=R^{yzy}_{BCD}=0.09, \\\\R^{zxx}_{BCD} = R^{zyy}_{BCD}=-0.18.
    \end{array}
\end{equation}
Therefore, remarkably, we see that in spite of having all its spectral weight localized in a delta function Drude-like peak, in TaAs the BCD contributes to the net rectification tensor with a magnitude that is comparable with the net contribution from shift currents, and sometimes they have oppposite signs as it is the case for the $R^{zxx}=R^{zyy}$.
The SRV for TaAs has a single non-trivial component $\boldsymbol{\eta}_{\mathrm{TaAs}} = (0,0,-0.015)$, which is oriented along its polar axes as depicted in Fig.[\ref{fig:sori}](a).

On the other hand, the lower Cc space group symmetry of LiAsSe$_2$ allows for a larger set of independent components:
\begin{equation}
\begin{array}{c}
R^{xxx}=0.12,\qquad R^{xyy}=-0.11, \qquad R^{xzz}=-0.18,\vspace{1mm}\\
    R^{zxx}=0.02,\qquad R^{zyy}=-0.09, \qquad R^{zzz}=0.36,\vspace{1mm}\\
    R^{xxz} = R^{xzx}=-0.005, \qquad R^{yyx}= R^{yxy}=-0.09,\vspace{1mm}\\
    R^{yyz} = R^{yzy}=-0.05,\qquad R^{zzx}=R^{zxz}=-0.16.
    \end{array}
\end{equation}
The SRV has two components $\boldsymbol{\eta}_{\mathrm{LiAsSe_2}} = (-0.033,0,0.068)$, with the higher SRV magnitude $\eta = 0.08$, or in other words, it produces a photocurrent that is is 8$\%$ of the ideal limit for p-n junction solar cells.  

We would like to mention in passing that the formulae for rectification conductivities of Refs.\cite{PhysRevB.23.5590,PhysRevB.19.1548} are incomplete and are missing terms even for the inter-band shift currents. Details of the discrepancy and comparison with other various formulae employed in the literature are presented in (see S.I.\ref{papermapp}). For the complete correct formulae for the general second order conductivity of any band structure see Ref.\cite{PhysRevLett.123.246602}.

\textit{\color{blue}Discussion}. We have introduced an operational definition of the Berry Phase Rectification Tensor as the second order polarization response of a material to an ideal delta function pulse of applied electric field. In the case of a time reversal invariant band structure this tensor is an intrinsic property of the material which only depends on the Berry connections of the Bloch bands and not on their energies. This tensor contains intra-band contributions which arise from the Berry curvature dipole and inter-band contributions which arise from the shift currents. One of the importances of this operational definition of the tensor is that it is in principle applicable in the presence of electron-electron interactions, which is an interesting direction for future research. We have computed these tensors for TaAs, where we find that remarkably the Berry Dipole contributions are comparable to the shift current contributions. 

We have also introduced a more technologically relevant figure of merit for bulk photo-current generation - the Solar Rectification Vector, which measures the net rectified current in response to ideal incident black-body radiation, in analogy to the model of solar cells in the classic work of Shockley and Queisser~\cite{ShockleyQueisser}. Using this, we have found that LiAsSe$_2$ can produce a large bulk current of about 0.1 electron per every incident photon. We have also generalized the formula of photo-current production defining the Glass coefficient \cite{Glass,PhysRevLett.109.116601} to include the black-body spectral distribution of intensity, the reflection coefficient at the surface of the material and the directional dependence of the linear conductivity of the material, which should provide accurate estimates of realistic photo-current production in solar cells based on bulk rectification mechanisms.

\textit{\color{blue}Acknowledgements}. We thank Qian Niu for pointing to us the operational
interpretation of the QRSR in terms of electric field pulses. We also thank Qian Niu, Cong Xiao, Jorge Facio, Jeroen van-den-Brink, Dennis Wawrzik, and Jhih-Shih You for valuable discussions.

\section{Supplementary Information}
\subsection{Sum rule computations}\label{SumRuleComp}
The rectification conductivity is:
\begin{multline}\label{fullcondM}
    \sigma^{\gamma\beta\alpha}_{(2)}(-\omega,\omega) =\sigma^{\gamma\beta\alpha}_{\mathrm{J}}(-\omega,\omega)+\sigma^{\gamma\beta\alpha}_{\mathrm{BCD}}(-\omega,\omega)+\\+\sigma^{\gamma\beta\alpha}_{\mathrm{I}}(-\omega,\omega)+\sigma^{\gamma\beta\alpha}_{\mathrm{S}}(-\omega,\omega),
\end{multline}
where J stands for ``Jerk", BCD for ``Berry curvature dipole", I for ``Injection" and S for ``shift current" respectively. We consider band structure with small relaxations ($\forall n,m,n\neq m : \Gamma \ll \epsilon_{n}-\epsilon_m$). Each contribution is given by:
\begin{equation}\label{term1}
    \sigma^{\gamma\beta\alpha}_{\mathrm{J}}(-\omega,\omega) = \frac{e^3}{\hbar^2}\int\frac{d\mathbf{k}}{(2\pi)^3}\sum_{nm}\frac{\frac{\partial \epsilon_n}{\partial k^\gamma}\frac{\partial^2 }{\partial k^\alpha\partial k^\beta}f_n\delta_{nm}}{\omega^2+\Gamma^2},
\end{equation}    
\begin{multline}    
    \sigma^{\gamma\beta\alpha}_{\mathrm{BCD}}(-\omega,\omega) =-\frac12\frac{e^3}{\hbar^2}\frac{1}{\omega+i\Gamma}\int\frac{d\mathbf{k}}{(2\pi)^3}\times\\\times\sum_{nm}\hat{A}^{\gamma}_{mn}\hat{A}^{\alpha}_{nm}\frac{\partial}{\partial k^\beta}(f_{m}-f_n)+\\+\left(\begin{array}{c}\alpha\leftrightarrow \beta\\\omega\leftrightarrow-\omega\end{array}\right).
\end{multline}
\begin{multline}
    \sigma^{\gamma\beta\alpha}_{\mathrm{I}}(-\omega,\omega) = \frac{e^3}{\hbar^2}\int\frac{d\mathbf{k}}{(2\pi)^3}\times\\\times\sum_{nm}\frac{(f_{m}-f_n)\hat{A}^{\beta}_{nm}\hat{A}^{\alpha}_{mn}(\frac{\partial}{\partial k^\gamma}\epsilon_n-\frac{\partial}{\partial k^\gamma}\epsilon_m)}{(\omega-\epsilon_n+\epsilon_m)^2+\Gamma^2}+\\+\left(\begin{array}{c}\alpha\leftrightarrow \beta\\\omega\leftrightarrow-\omega\end{array}\right),
\end{multline}
\begin{multline}\label{termlast}
    \sigma^{\gamma\beta\alpha}_{\mathrm{S}}(-\omega,\omega) = \frac12\frac{e^3}{\hbar^2}\int\frac{d\mathbf{k}}{(2\pi)^3}\times\\\times\sum_{nm}\Bigg\{ \hat{A}^{\gamma}_{mn}\frac{\partial}{\partial k^\alpha}\frac{(f_{n}-f_m)\hat{A}^{\beta}_{nm}}{\omega-\epsilon_{n}+\epsilon_{m}+i\Gamma} 
    +\\+ i\frac{(f_{n}-f_m)\hat{A}^{\beta}_{nm}}{\omega  -\epsilon_{n}+\epsilon_{m}+i\Gamma}\sum_c \bigg[\hat{A}^\alpha_{mc}\hat{A}^{\gamma}_{cn}-\hat{A}^{\gamma}_{mc}\hat{A}^\alpha_{cn}\bigg]\Bigg\}+\\+\left(\begin{array}{c}\alpha\leftrightarrow \beta\\\omega\leftrightarrow-\omega\end{array}\right).
\end{multline}

For purposes of ensemble averaging over incoming random light, we need to consider only diagonal components of the nonlinear rectification conductivity, namely $\sigma_{(2)}^{\gamma\alpha\alpha}$.

To simplify the notation for further analysis we write:
\begin{equation}
    Q^{\gamma\beta\alpha}_{nm} = \hat{A}^{\beta}_{nm}  \frac{\partial}{\partial k^\alpha}\hat{A}^{\gamma}_{mn}-i\hat{A}^{\beta}_{nm}\sum_c \left[\hat{A}^\alpha_{mc}\hat{A}^{\gamma}_{cn}-\hat{A}^{\gamma}_{mc}\hat{A}^\alpha_{cn}\right].
\end{equation}

Thus, the real part of the conductivity which is the one that enters into the SRV, is:
\begin{multline}
    \mathrm{Re}\left[\sigma^{\gamma\alpha\alpha}_{\mathrm{J}}(-\omega,\omega)\right] =\\= \frac{e^3}{\hbar^2}\int\frac{d\mathbf{k}}{(2\pi)^3}\sum_{nm}\frac{\frac{\partial \epsilon_n}{\partial k^\gamma}\frac{\partial^2 }{\partial k^\alpha\partial k^\alpha}f_n\delta_{nm}}{\omega^2+\Gamma^2},
\end{multline}    
\begin{multline}    
    \mathrm{Re}\left[\sigma^{\gamma\alpha\alpha}_{\mathrm{BCD}}(-\omega,\omega)\right] =\frac{1}{2}\frac{e^3}{\hbar^2}\frac{\Gamma}{\omega^2+\Gamma^2}\int\frac{d\mathbf{k}}{(2\pi)^3}\times\\\times\sum_{nm}f_{m}\frac{\partial}{\partial k^\beta}\Omega_m^{\alpha\gamma},
\end{multline}
\begin{multline}
     \mathrm{Re}\left[\sigma^{\gamma\alpha\alpha}_{\mathrm{I}}(-\omega,\omega)\right] = \frac{e^3}{\hbar^2}\int\frac{d\mathbf{k}}{(2\pi)^3}\times\\\times\sum_{nm}\frac{(f_{m}-f_n)\hat{A}^{\alpha}_{nm}\hat{A}^{\alpha}_{mn}(\frac{\partial}{\partial k^\gamma}\epsilon_n-\frac{\partial}{\partial k^k}\epsilon_m)}{(\omega-\epsilon_n+\epsilon_m)^2+\Gamma^2},
\end{multline}
\begin{multline}
    \mathrm{Re}\left[\sigma^{\gamma\alpha\alpha}_{\mathrm{S}}(-\omega,\omega)\right] = \frac{e^3}{\hbar^2}\int\frac{d\mathbf{k}}{(2\pi)^3}\sum_{nm}(f_{m}-f_n)\times\\\times\frac{\mathrm{Re}\left[Q^{\gamma\alpha\alpha}_{nm}\right](\omega-\epsilon_{n}+\epsilon_{m})+\Gamma\mathrm{Im}\left[Q^{\gamma\alpha\alpha}_{nm}\right]}{(\omega-\epsilon_{n}+\epsilon_{m})^2+\Gamma^2}.
\end{multline}

Time reversal symmetry (TRS) implies:
\begin{gather}
    \mathcal{T}\mathbf{A}_{mn}(k)\mathcal{T}^{-1} = \mathbf{A}_{nm}(-k),\\
    \mathcal{T}\mathbf{r}_{mn}(k)\mathcal{T}^{-1} = \mathbf{r}_{nm}(-k),\\\mathcal{T}\mathbf{v}_{mn}(k)\mathcal{T}^{-1} = -\mathbf{v}_{nm}(-k),
\end{gather}

TRS makes the Jerk, Injection and the part proportional to $\mathrm{Re}\left[Q^{\gamma\alpha\alpha}\right]$ of the Shift current vanish after momentum integration. Thus the contribution to the total rectification conductivity is given by BCD and resonant part of the shift current:
\begin{multline}    
    \mathrm{Re}\left[\sigma^{\gamma\alpha\alpha}_{\mathrm{BCD}}(-\omega,\omega)+\sigma^{\gamma\alpha\alpha}_{\mathrm{S}}(-\omega,\omega)\right] =\frac{e^3}{\hbar^2}\int\frac{d\mathbf{k}}{(2\pi)^3}\times\\\Bigg\{\frac{\Gamma}{\omega^2+\Gamma^2}\sum_{m}f_{m}\frac{\partial}{\partial k^\alpha}\Omega_m^{\alpha\gamma} +\\+\sum_{nm}(f_{m}-f_n)\frac{\Gamma\mathrm{Im}\left[Q^{\gamma\alpha\alpha}_{nm}\right]}{(\omega-\epsilon_{n}+\epsilon_{m})^2+\Gamma^2}\Bigg\}.
\end{multline}

In $\Gamma\rightarrow 0$ limit, we can simplify the expression using the following relation:
\begin{equation}
    \lim_{a\rightarrow0^+}\frac{a}{d_z^2+a^2}=\pi \delta(d_z).
\end{equation}

Now we can write a general version of the sum rule by including the spectral weight, which we assume to be a smooth function of $\omega$, to obtain:
\begin{multline}\label{current3}
    \int d\omega F(\omega)\sigma^{\gamma\alpha\alpha}(-\omega,\omega)=\pi\frac{e^3}{\hbar^2}\sum_{\alpha,nm}f_{nm}F(|\epsilon_{nm}|)\times\\\times\Bigg\{A^{\alpha}_{nm}i\partial^\alpha A^\gamma_{mn}+\sum_l A^{\alpha}_{nm}\left[A^{\alpha}_{ml}A^{\mu}_{ln}-A^{\gamma}_{ml}A^{\alpha}_{ln}\right]\Bigg\},
\end{multline}
where $f_{nm} = f_n - f_m$ and all zero-frequency contributions vanish under the assumption of absence of spectral weight ($F(\omega\rightarrow 0) = 0$).

Now in the case of the operational definition of the BPRT, we consider the net change of the polarization coursed by pulse-like Electric field:
\begin{gather}
    \mathbf{E}(t) = \mathbf{E}_0\tau_0\delta(t),\qquad
    \mathbf{E}(\omega) = \frac{\mathbf{E}_0\tau_0}{2\pi},\\
    \Delta \mathbf{P} = \int_0^\infty \mathbf{j}(t)dt.
\end{gather}

In this set-up, without assuming time reversal symmetry the resulting change of the polarization is:
\begin{multline}
    \Delta P^\gamma = \frac{E_0^\alpha E_0^\beta}{4}\tau_0^2\frac{e^3}{\hbar^2}\times\\\times\Bigg\{
    \frac{1}{\Gamma}\left(\left\langle\partial^\alpha\partial^\beta v^{\gamma}\right\rangle+\sum_{ab}f_{ab}A^\beta_{ab}A^{\alpha}_{ba}\Delta^\gamma_{ba}\right)+\\+\left\langle\left[A^{\beta},i\partial^{\alpha}A^{\gamma}\right]\right\rangle+\left\langle\partial^\beta\Omega^{\alpha\mu}\right\rangle+\\+\left\langle\left[A^\beta,\left[A^\alpha,\bar{A}^\gamma\right]\right]\right\rangle+\left(\alpha\leftrightarrow \beta\right)\Bigg\},
\end{multline}
where $\sum_{ab} = \sum_{ab}\int d \mathbf{k}/ (2\pi)^d$. As we see this result contains two extra terms proportional to $1/\Gamma$ which arise from the Jerk and injection current terms, but which vanish under the assumption of time-reversal symmetry.

\subsection{Black body radiation}\label{BBrad}
The quantization of EM field in Coulomb gauge can be performed by solving the vector potential wave equation in a box (volume $V$) with periodic boundary conditions:
\begin{equation}
    \frac{\partial^2 \mathbf{A}}{\partial t^2} - c^2\Delta \mathbf{A} = 0.
\end{equation}

The solution is given by:
\begin{multline}
    \mathbf{A}(\mathbf{r},t) = \sum_{\mu,\mathbf{k}}\sqrt{\frac{\hbar}{2\epsilon_0V\omega_{\mathbf{k}}}}\bigg[a_{\mathbf{k}}^\mu e^{-i(\omega_{\mathbf{k}}t-\mathbf{k}\cdot\mathbf{r})}+\\+a^{\dagger\mu}_{\mathbf{k}} e^{i(\omega_{\mathbf{k}}t-\mathbf{k}\cdot\mathbf{r})}\bigg]\mathbf{e}_\mu(\mathbf{k}),
\end{multline}
where $\omega_{\mathbf{k}}=c|\mathbf{k}|$ and $V$ is the box volume. The electric field is $\mathbf{E}=-\partial \mathbf{A}/\partial t$:
\begin{multline}
    \mathbf{E}(\mathbf{r},t) = i \sum_{\mu,k}\sqrt{\frac{ \hbar\omega_\mathbf{k}}{2\epsilon_0V}}\bigg[a_{\mathbf{k}}^\mu e^{-i(\omega_{\mathbf{k}}t-\mathbf{k}\cdot\mathbf{r})}-\\-a^{\dagger\mu}_{\mathbf{k}} e^{i(\omega_{\mathbf{k}}t-\mathbf{k}\cdot\mathbf{r})}\bigg]\mathbf{e}_\mu(\mathbf{k}),
\end{multline}
where $\mathbf{e}^{\pm}=\frac{\mp1}{\sqrt{2}}(\mathbf{e}_x\pm i\mathbf{e}_y), \mathbf{e}_x\cdot\mathbf{k}=\mathbf{e}_y\cdot\mathbf{k}=\mathbf{e}_x\cdot\mathbf{e}_y=0$. The Hamiltonian for each mode is $\mathcal{H}_m=\hbar\omega_m(a^\dagger_m a_m +\frac12)$.
Using the following relations:
\begin{gather}
    \left[a_\mathbf{k}^{\mu},a_{\mathbf{k}^\prime}^{\mu^\prime}\right] = 0,\qquad
    \left[a_\mathbf{k}^{\mu\dagger},a_{\mathbf{k}^\prime}^{\mu^\prime\dagger}\right] = 0,\\
    \left[a_\mathbf{k}^{\mu},a_{\mathbf{k}^\prime}^{\mu^\prime\dagger}\right] = \delta_{\mathbf{k}\mathbf{k}^\prime}\delta_{\mu\mu^\prime},\qquad
    \langle a_\mathbf{k}^{\mu},a_{\mathbf{k}^\prime}^{\mu^\prime\dagger}\rangle =\delta_{\mathbf{k}\mathbf{k}^\prime}\delta_{\mu\mu^\prime} n_{\mathbf{k}} ,\\n_{\mathbf{k}} = \frac{1}{e^{\frac{\hbar c{k}}{k_\mathbf{B}T}}-1},\qquad \sum_{\mu}\mathbf{e}^\alpha_\mu(\mathbf{k})\mathbf{e}^{\beta*}_{\mu}(\mathbf{k})=\delta_{\alpha\beta}-\hat{\mathbf{k}}_{\alpha}\hat{\mathbf{k}}_{\beta},
\end{gather}
one can obtain the electric field auto-correlation function:
\begin{multline}\label{corr2}
    \langle {E}^\alpha(\mathbf{r},t){E}^{\beta}(\mathbf{r},t^\prime) \rangle = \\=\frac{\hbar}{\epsilon_0V}\sum_{\mathbf{k}}(\delta_{\alpha\beta}-\hat{\mathbf{k}}_{\alpha}\hat{\mathbf{k}}_{\beta})\omega_\mathbf{k}n_\mathbf{k}\cos[\omega_\mathbf{k}(t-t^\prime)],
\end{multline}
where $\hat{\mathbf{k}}^\alpha$ is the unit vector of momentum, and $\sum_{\mathbf{k}}=V\int d\mathbf{k}/(2\pi)^3$. In the above formula all modes of light contribute to produce a fully isotropic correlator of electric fields at a given point {\bf r}. However, when the radiation that arrives at {\bf r} is highly directional we can modify the formula above by a one with an extra function inside the momentum integral $f(\hat{{\bf k}})$:
\begin{multline}\label{corr02}
    \langle {E}^\alpha(\mathbf{r},t){E}^{\beta}(\mathbf{r},t^\prime) \rangle = \\=\frac{\hbar}{\epsilon_0V}\sum_{\mathbf{k}}(\delta_{\alpha\beta}-\hat{\mathbf{k}}_{\alpha}\hat{\mathbf{k}}_{\beta})\omega_\mathbf{k}n_\mathbf{k}\cos[\omega_\mathbf{k}(t-t^\prime)]f(\hat{\mathbf{k}}),
\end{multline}
which phenomenologically takes into account the preferred directionality of the incoming light at a given point {\bf r}. In the special case in which the function $f(\hat{{\bf k}})=1$ only for a very small region of solid angles,  $\Delta \Omega$, around a specific direction $\hat{{\bf k}}=\hat{{\bf k}}_0$, we have that:
\begin{multline}\label{corr3}
    \langle {E}^\alpha(\omega_1){E}^{\beta}(\omega_2) \rangle = \\=\frac{\Delta\Omega}{4\pi}\frac{\hbar(\delta_{\alpha\beta}-\hat{\mathbf{k}}_0^\alpha\hat{\mathbf{k}}_0^\beta)}{(2\pi)^2\epsilon_0(\beta\hbar c)^3}\frac{(\hbar\beta|\omega_1|)^3}{e^{\hbar\beta|\omega_1|}-1}\delta(\omega_1+\omega_2).
\end{multline}

In the case of the light emitted from the sun and arriving on earth, $\Delta\Omega= 4\pi (R_\mathbf{E}/R_\mathbf{S})^2$, $\hat{\bf k}_0$ is the unit vector defining the direction from sun to earth, $R_\mathbf{S}$ is the radius of the sun, and $R_\mathbf{E}$ is the distance of the sun to the earth.

The formula above is expected to describe the ideal solar radiation before entering the earth atmosphere, but the solar irradiance spectrum for practical applications on the surface of earth can still be reasonably approximated by it (see e.g.~\cite{metrodata}). Upon entering the atmosphere the spectrum intensity gets reduced at specific frequencies corresponding to absorption of molecules in the atmosphere. There will be also a randomization of the transversality of the electric field due to elastic scattering. This could be accounted for by averaging the projector over some distribution of $\hat{\bf k}_0$, but these subtle details are beyond the scope of our discussion. In the main text we have omitted the transverse projector and replaced it with a usual delta function with all components for simplicity. In practice as along as the material is oriented so that its SRV is along the plane that is orthogonal to $\hat{{\bf k}}_0$ one obtains the same current from Eq.(\ref{corelfun}) from the as the one one would obtain from the Eq.(\ref{corr3}) that takes into account the transverse nature of incident light. Notice, however, that the transverse nature of incident light has explicitly been taken into account in derivation of the more microscopically accurate formula in Eq.(\ref{igamma}) of the main text.

Let us now compute the photon flux on earth, namely the number of incident photons per unit area per unit time.
The photon density, $n$, and current density, $\mathbf{F}^\mu$, are given by:
\begin{gather}
    a^\mu(\mathbf{r},t) = \frac{1}{\sqrt{V}}\sum_{\mathbf{k}}a^\mu_{\mathbf{k}}e^{-i\omega_{\mathbf{k}}t+i\mathbf{k}\cdot\mathbf{r}},\\
    n^\mu(\mathbf{r}) = a^{\mu\dagger}(\mathbf{r})a^\mu{(\mathbf{r})},\\
    \mathbf{F}^\mu(\mathbf{r},t) = \sum_{\mathbf{q}}\mathbf{F}^\mu_\mathbf{q} e^{-i\omega_{\mathbf{q}}t+i\mathbf{q}\cdot\mathbf{r}},
\end{gather}
where $\mu$ labels the polarization state and space indices are implicit. The current can be obtained by using the continuity equation for photon number:
\begin{equation}
    \partial_tn^\mu = -\mathbf{\nabla}\cdot\mathbf{F^\mu}.
\end{equation}

Which leads to the following expression for the photon current per unit of area per unit of time:
\begin{gather}
    F_x = \sum_\mu \mathbf{F}^\mu \cdot \hat{\bf x} = \frac{c}{V}\sum_{\mu,\mathbf{k},k_x>0}\left(\hat{\mathbf{k}}\cdot\mathbf{x}\right)a^{\dagger\mu}_\mathbf{k}a^\mu_{\mathbf{k}}.
\end{gather}

The expression above would give us the number of photons moving on a single direction (e.g. towards the rigth) on a given surface of a black-body, which can be taken to be the surface of the Sun.
On the surface of the Earth the photon flux is given by:
\begin{multline}\label{corr5}
    F_s = \left(\frac{R_\mathbf{S}}{R_{\mathbf{E}}}\right)^2\frac{c}{V}\sum_{\mu,\mathbf{k},k_x>0}\left(\hat{\mathbf{k}}\cdot\mathbf{x}\right)\left\langle a^{\dagger\mu}_\mathbf{k}a^\mu_{\mathbf{k}}\right\rangle=\\=\left(\frac{R_\mathbf{S}}{R_{\mathbf{E}}}\right)^2\frac{2\zeta(3)}{(2\pi)^2}\frac{c}{(\beta\hbar c)^3}.
\end{multline}

Then by combining Eqs.(\ref{corr3}) and Eq.(\ref{corr5}) with the definition from Eq.(\ref{SQsm}) of the main text, one arrives at the result of the Solar Rectification Vector, shown in Eq.(\ref{etagamma}) of the main text. 

On the other hand, the number of photons that have energy above the gap of the material that can be used to defined the fraction of absorbed photons in Eq.(14) of the main text is given by:
\begin{equation}
    F(\Delta_0)=\left(\frac{R_\mathbf{S}}{R_{\mathbf{E}}}\right)^2\frac{c}{V}\sum_{\mu,\mathbf{k}\atop  k_x>0}\Theta(ck-\Delta_0)\left(\hat{\mathbf{k}}\cdot\mathbf{x}\right)\left\langle a^{\dagger\mu}_\mathbf{k}a^\mu_{\mathbf{k}}\right\rangle,
\end{equation}
where $\Delta_0$ is the band-gap of the material, $\Theta$ is the heaviside function.  The ratio of absorbed photons, $r$, in Eq.(\ref{jsq}) of the main text, can thus be obtained as $r=F(\Delta_0)/F_s$.

\subsection{Material penetration}\label{GlassSupp}

\begin{figure}[t]
    \centering
    \includegraphics[width=0.48\textwidth]{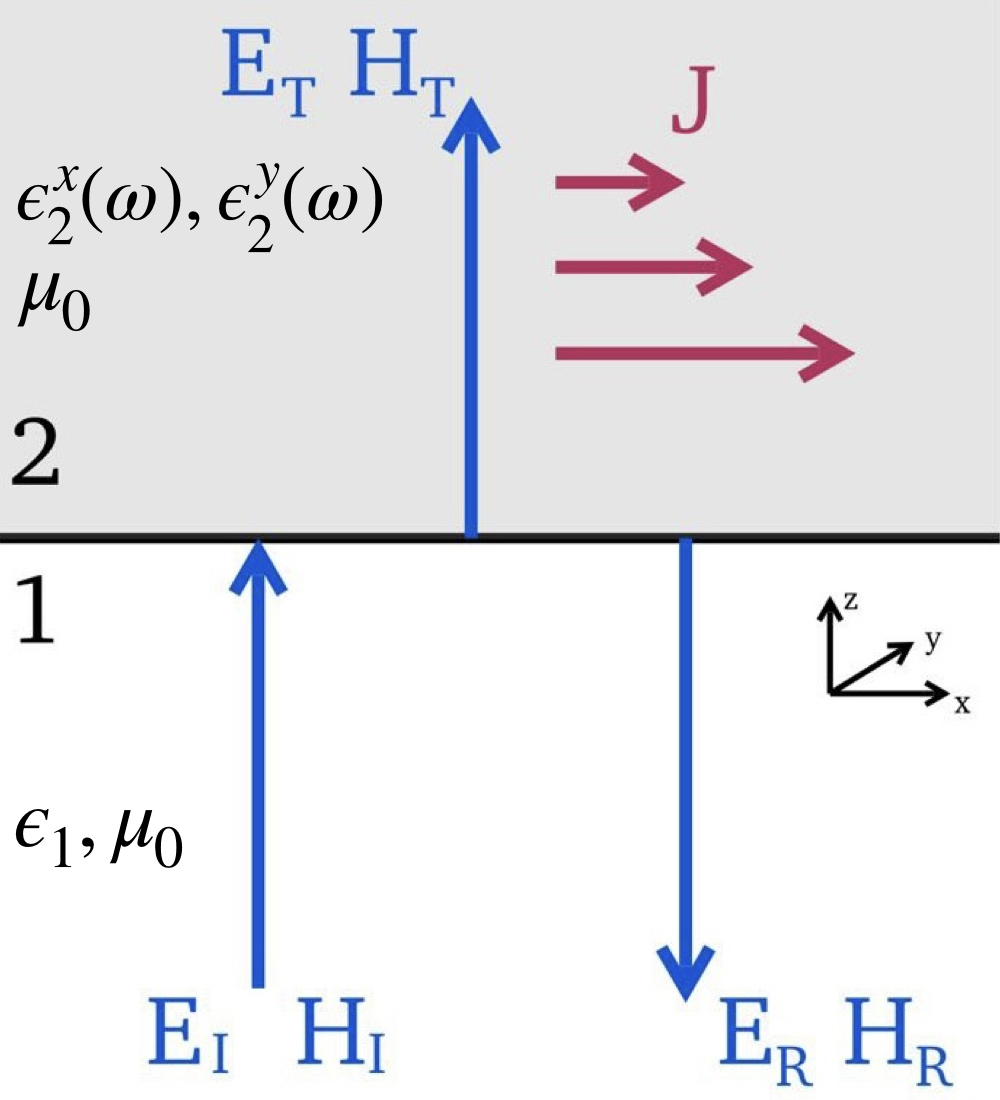}
    \caption{Illustration of the geometry for incident (I), transmited (T) and reflected (R) light at the solar cell. For simplicity, we considered the light to be perpendicular to the surface. At the interface (solid black line) between vacuum (region 1) and the bulk photovoltaic material that makes the solar cell (region 2) the incoming light (I) splits onto reflected (R) and transmitted (T). The transmitted component generates the current J. $\mu_0$ is magnetic permitivity and $\epsilon_1$, $\epsilon^{x,y}_2$ are permittivities of vacuum and the bulk photo-voltaic material respectively.}
    \label{fig:specw8}
\end{figure}

In this section we take into account that the intensity of electric field decays inside the material, and derive the expression for the total current. For simplicity, we assume that incident radiation is normal to the solar cell surface so that that the electric field is parallel to the surface, namely along the XY plane in Fig.[\ref{fig:specw8}], and that the induced photo-current also flows along this XY plane.
To estimate the electric field penetration, we use linear response theory. Thus from Maxwell equations we have:
\begin{gather}
    \mathbf{E}_I(\mathbf{z},t) = \mathbf{E}_Ie^{i(k_1{z}-\omega t)},\quad \mathbf{E}_R(\mathbf{z},t) = \mathbf{E}_Re^{-i(k_1{z}+\omega t)},\\ \mathbf{H}_I(\mathbf{z},t) = \mathbf{H}_Ie^{i({k}_1{z}-\omega t)},\quad \mathbf{H}_R(\mathbf{z},t) = \mathbf{H}_Re^{-i({k}_1{z}+\omega t)},\\
    \mathbf{E}_{I,R} = \left(E^x_{I,R},E^y_{I,R},0\right),\quad \mathbf{H}_{I,R} = \left(H^x_{I,R},H^y_{I,R},0\right),\\
        \mathbf{E}_T(\mathbf{z},t) = \mathbf{E}_{T1}e^{i(k_{2y}{z}-\omega t)}+\mathbf{E}_{T2}e^{i(k_{2x}{z}-\omega t)},\\
    \bm{\nabla}\times\mathbf{H}=-i\mathbf{J}-\epsilon_0\omega\mathbf{E},\\
    \bm{\nabla}\times\mathbf{E}=\omega\mu_0\mathbf{H},\qquad J^\alpha(\omega) =\sigma^{\alpha\beta}_{(1)}(\omega)E^{\beta}(\omega).
\end{gather}

For simplicity we assume the magnetic permitivities of the vacuum and the bulk photovoltaic material to be frequency independent constants. We also assume that the material has a point group so that its linear conductivity is diagonal in its set of principal axes, and we take its z-axis to coincide with the direction of incoming light. The frequency dependence of these diagonal components of the linear conductivity of the material can be then obtained from the microscopic band-structure as follows:
\begin{multline}
          \sigma^{\beta\beta}_{(1)}(\omega)=-\frac{e^2}{\hbar}\sum_{nm}\Bigg\{\delta_{nm}\partial^\beta\epsilon_{n}\frac{i\partial^\beta f_n}{\omega+i\Gamma}+\\+i(f_{n}-f_{m})(\epsilon_{n}-\epsilon_{m})\frac{A^{\beta}_{nm}A^\beta_{mn}}{\omega-\epsilon_{nm}+i\Gamma}\Bigg\}.  
\end{multline}
Where in this expression the repeated indices are not summed over. From the above one can obtain the dispersion relations of light outside and inside the material to be:
\begin{gather}
    k_1^2=\mu_0\omega^2\epsilon_0,\quad k_{2x}^2=\mu_0\omega^2\epsilon^x_2(\omega),\quad k_{2y}^2=\mu_0\omega^2\epsilon^y_2(\omega),\\  \epsilon^x_2 (\omega)= \epsilon_0+i\frac{\sigma^{xx}_{(1)}(\omega)}{\omega},\quad
    \epsilon_2^y (\omega)= \epsilon_0+i\frac{\sigma^{yy}_{(1)}(\omega)}{\omega}.
\end{gather}

By taking into account standard boundary conditions Ref.\cite{zangwill_2012}, we can obtain the following relation describing the transmitted, T, component of the electric field inside the material for a given incident component, I:
\begin{multline}
    \mathbf{E}_T(\omega,z) =\\=\frac{2 n_1}{n_1+n^x_2(\omega)}(E^x_I(\omega),0,0) e^{i(\mathrm{Re}\left[{k}_{2x}\right]{z}-\omega t)}e^{-\mathrm{Im}\left[{k}_{2x}\right]{z}}+\\+\frac{2 n_1}{n_1+n^y_2(\omega)}(0,E^y_I(\omega),0) e^{i(\mathrm{Re}\left[{k}_{2y}\right]{z}-\omega t)}e^{-\mathrm{Im}\left[{k}_{2y}\right]{z}},
\end{multline}
where $n_1 =c\sqrt{\mu_0\epsilon_0},~ n^\beta_i = c\sqrt{\mu_0\epsilon^\beta_i(\omega)},~ k_{i\beta} = \omega n^\beta_i(\omega)/c$. Thus the electric field correlation in bulk is given by:
\begin{multline}
    \left\langle E^\alpha_T(\omega,z)E^\beta_T(-\omega,z) \right\rangle=\\=\delta_{\alpha\beta}\left|\frac{2n_1}{n_1+n^\beta_2(\omega)}\right|^2 e^{-2\frac{\omega}{c}\mathrm{Im}[n^\beta_2(\omega)]z}\left\langle E^\beta_I(\omega)E^\beta_I(-\omega) \right\rangle,
\end{multline}
where in the above expression the repeated indices are not being summed over and indices are understood to take only $(x,y)$ components. Thus the total current is:
\begin{gather}
    \frac{\overline{I^\gamma}}{W} =\int_{-\infty}^{\infty}d\omega |T^\beta(\omega)|^2\frac{\sigma_{(2)}^{\gamma\beta\beta}(-\omega,\omega)}{\alpha^\beta(\omega)}I^{\beta\beta}(\omega),\\
    T^\beta(\omega) =\frac{2n_1}{n_1+n^\beta_2(\omega)},\quad \alpha^\beta(\omega)=2\frac{\omega}{c}\mathrm{Im}[n^\beta_2(\omega)],
\end{gather}
where $W$ is the width of the sample, and the repeated index $\beta$ is summed over the two possible components $(x,y)$.  
\begin{figure*}
\centering
\includegraphics[width=0.98\textwidth]{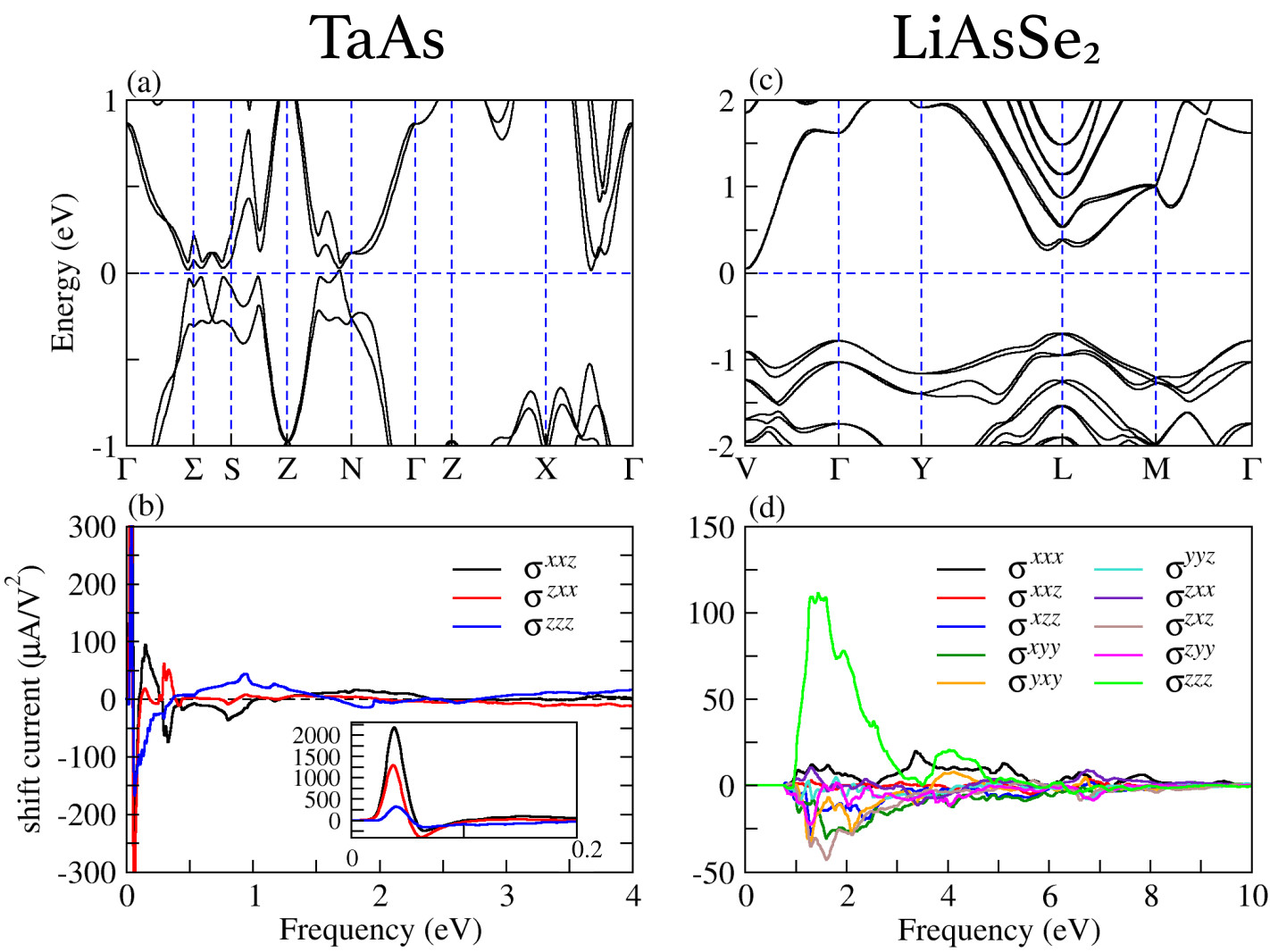}
\caption{Band structures of (a) TaAs and (b) LiAsSe$_2$. Rectification conductivities of (c) TaAs and (d) LiAsSe$_2$.}
\label{fig:fig5}
\end{figure*}

\subsection{Computation details}\label{DFTdetails}
The exchange and correlation energies were considered in the generalized gradient approximation with Perdew-Burke-Ernzerhof functional \cite{PhysRevLett.77.3865}. We projected the Bloch wave functions into high-symmetry atomic-orbital-like Wannier functions with diagonal position operator and considered all the electrons, which guarantees the full band overlap from ab−initio and Wannier functions in the energy window from -20 to 20 eV. The calculated band structures of TaAs and LiAsSe2 are shown in Fig[\ref{fig:fig5}]. Based on the highly symmetric Wannier functions, we constructed tight-binding (TB) model Hamiltonians and calculated the Berry curvature dipole and second order optical conductivity tensor. The $\gamma$ (with $\alpha, \beta, \gamma = x, y, z$) component of Berry curvature for the n−th band at k
point is computed by following:
\begin{equation}
    \Omega_{n}^\gamma(\mathbf{k}) = \frac{\epsilon^{\gamma\alpha\beta}}{2}\left(\frac{\partial \hat{A}_{n}^{\beta}}{\partial k^\alpha}-\frac{\partial \hat{A}_{n}^{\alpha}}{\partial k^\beta}\right),    
\end{equation}
where, $\hat{A}_{n}^{\beta}=\bra{n,\mathbf{k}}i\partial ^\beta\ket{n,\mathbf{k}}$ is the Berry connection of $n$-th band of Hamiltonian $H(\mathbf{k})$. The Berry curvature dipole was calculated from
the Berry curvature in a differential manner and with integral in the whole $\mathbf{k}$-space \cite{PhysRevLett.115.216806}:

\begin{equation}
    D_{\alpha\beta} = \int \frac{d\mathbf{k}}{(2\pi)^3}\sum_n f_n\frac{\partial\Omega^\beta(\mathbf{k})}{\partial k^\alpha}=\left\langle\frac{\partial\Omega^\beta(\mathbf{k})}{\partial k^\alpha}\right\rangle,
\end{equation}
where $\Omega^\delta_n= \varepsilon^{\delta\alpha\beta}\partial A^\beta/\partial k^\alpha$ is the Berry curvature vector and $f_n$ is a Fermi-Dirac distribution. The relation between two and three index BCD tensors is:
\begin{equation}
    \varepsilon^{\alpha\gamma\delta}D^{\beta\delta} = \partial^\beta \Omega^{\alpha\gamma}.
\end{equation}

The second order optical conductivity tensor is calculated by the following expression:
\begin{multline}
    \sigma^{\gamma\alpha\beta}_{(2)}(-\omega,\omega)=\frac{\pi e^3}{2\hbar^2}\int\frac{d\mathbf{k}}{(2\pi)^3}\sum_{nm}(f_n-f_m)\times\\\times\Big\{A^\beta_{nm}i\partial^\alpha A^\gamma_{mn}+A^\beta_{nm}\left[A^\alpha,\bar{A}^\gamma\right]_{mn} +\\+(\alpha\leftrightarrow\beta)\Big\}\delta\left(\omega-\frac{\epsilon_{n}-\epsilon_m}{\hbar}\right),
\end{multline}
here $\epsilon_n$ is n-th band energy. The expression above is equivalent to expressions discussed in \cite{PhysRevLett.123.246602,PhysRevB.61.5337}(see S.I. \ref{papermapp}). The numerical results are consistent with the symmetry analysis. The plots showing the frequency dependence of the second order rectification conductivities of TaAs and LiAsSe$_2$ are shown in Fig.[\ref{fig:fig5}]. These plots show only the interband shift currents, since the BCD is simply a delta function peak at zero frequency.

After a k-grid convergency test, it was found that the change of BCD for TaAs is less than 5$\%$ with k-grid increasing from $360\times360\times360$ to $480\times480\times480$. A similar k-grid convergency was obtained with a k-grid of increasing from $240\times240\times240$ to $360\times360\times360$ for the second order optical conductivity in TaAs and LiAsSe$_2$.
\subsection{Equivalence with other formalisms in the literature.}\label{papermapp}
Here we discuss the equivalence of our formulae for the rectification conductivity from Eq.(\ref{fullcondM}) and others employed in the literature\cite{PhysRevB.19.1548,PhysRevB.23.5590,PhysRevB.97.245143,PhysRevB.61.5337,PhysRevB.99.045121,PhysRevB.98.165113,2019arXiv190808878Z,PhysRevB.52.14636}. Importantly, we show that shift currents derived in classic papers Refs.\cite{PhysRevB.23.5590,PhysRevB.19.1548} are only valid for Galilean systems, where spin orbit coupling is neglected, but not in general.  
\subsubsection{\textbf{Comparison with} Refs.\cite{PhysRevB.61.5337,PhysRevB.97.245143}}
Here we will demonstrate that the interband rectification current in time reversal invariant materials arising from $\sigma_I + \sigma_S$ from Eqs(\ref{term1}-\ref{termlast}), is identical to that from Refs.\cite{PhysRevB.61.5337,PhysRevB.98.165113}. This is the part of the rectification conductivity that controls the inter-band contribution to the QRSR and BPRT, but it is missing the intraband BCD contribution. This interband part however is the only one contributing to the SRV.

We define the following notation:
\begin{gather}\label{defff}
    r^{a}_{nm}=\frac{v^a_{nm}}{i\epsilon_{nm}}=\bar{A}^a_{nm}, \qquad(n\neq m),\\
    r^{a;b}_{nm}=\partial^b r^a_{nm}-ir^a_{nm}(A^b_{nn}-A^b_{mm}),
\end{gather}
where $\epsilon_{nm}=\epsilon_n-\epsilon_m$ and $(a,b,c)$ are space indices. Alternatively, this can be written as:
\begin{multline}
    r^{a;b}_{nm}=\frac{i}{\epsilon_{nm}}\Bigg\{\frac{v^a_{nm}\Delta^b_{nm}}{\epsilon_{nm}}+\frac{v^b_{nm}\Delta^a_{nm}}{\epsilon_{nm}}-\vartheta^{ab}_{nm}+\\+\sum_{p\neq n,m}\left(\frac{v^a_{np}v^b_{pm}}{\epsilon_{pm}}-\frac{v^a_{pm}v^b_{np}}{\epsilon_{np}}\right)\Bigg\},
\end{multline}
where $\Delta^a_{nm}=v_{nn}^a-v_{mm}^a$, and
\begin{multline}
    \vartheta^{ab}_{nm}=\bra{n}\partial^a\partial^b H\ket{m}=\\=\partial^bv^a_{nm}+i\sum_p(v^a_{np}r^b_{pm}-r^b_{np}v^a_{pm}).
\end{multline}

By using the identity identity $[r^a,r^b]=0$, one obtains:
\begin{multline}
    \partial^a A^b_{nm}-i[A^a,\bar{A}^b]_{nm}=\\=\partial^bA^a_{nm}-iA^a_{nm}(A^b_{nn}-A^b_{mm}),
\end{multline}
which leads to an alternative form to denote the generalized derivative:
\begin{multline}\label{gender1}
        r^{a;b}_{nm}=\partial^b A^a_{nm}-iA^a_{nm}(A^b_{nn}-A^b_{mm})=\\=\partial^a A^b_{nm}-i\sum_p\left\{A^c_{np}\bar{A}^a_{pm}-\bar{A}^a_{np}A^c_{pm}\right\},
\end{multline}
for $n\neq m$. By using Eq.(\ref{defff}) and Eq.(\ref{gender1}) one can obtain that the conductivity from \cite{PhysRevB.97.245143,PhysRevB.61.5337} can be written as:
\begin{multline}
    \sigma^{abc}(0,\omega,-\omega)=-\frac{i\pi e^3}{2\hbar^2}\int\frac{d\mathbf{k}}{(2\pi)^3}\sum_{nm}f_{nm}\times\\\times\left(r^b_{mn}r^{c;a}_{nm}+(b \leftrightarrow c)\right)\delta(\frac{\epsilon_{mn}}{\hbar}-\omega).
\end{multline}
this is the same as the sum of the real part of injection and shift  currents $\sigma_{S} + \sigma_I$ assuming time-reversal invariant conditions from Eq.(\ref{fullcondM}), discussed in section (see S.I.\ref{SumRuleComp}), which were the ones employed in the main text. In the main text, we are also omitting the ``0" in the conductivity arguments, but it is implicit every time we refer to a ``rectification" conductivity. 

\subsubsection{\textbf{Comparison with} 
Refs.\cite{PhysRevB.23.5590,PhysRevB.19.1548,PhysRevB.99.045121}}

We will here compare our formula from the current paper and Ref.\cite{PhysRevLett.123.246602} to the formulae from the classic Ref.\cite{PhysRevB.23.5590,PhysRevB.19.1548} (see eg. Ref\cite{PhysRevB.97.241118,2019arXiv190808878Z} for an application of this) and those of more recent work of Ref.\cite{PhysRevB.99.045121} (see also Ref.\cite{conv} for closely related formulae to Ref.\cite{PhysRevB.99.045121}). We will show that Refs.\cite{PhysRevB.23.5590,PhysRevB.19.1548} are missing crucial corrections that can only be neglected if one assumes the system to have an underlying parabolic dispersion (namely ignores spin-orbit coupling). On the other hand we will show that, upon proper regularization of infinitesimal imaginary parts in the frequency denominators, the formulae of Ref.\cite{PhysRevB.99.045121} are equivalent to those we employ in the current paper and in our previous work Ref.\cite{PhysRevLett.123.246602}.

Following Ref.\cite{PhysRevB.99.045121} we write the second order perturbation to the Hamiltonian, as:
\begin{gather}
    \hat{H} = \hat{H}_0 + \hat{V}(t),\label{PMsetup1}\\
    \hat{V}(t)=\frac{e}{\hbar}A^\alpha \hat{h}^\alpha+\frac12\left(\frac{e}{\hbar}\right)^2A^{\alpha}(t)A^{\beta}(t)\hat{h}^{\alpha\beta}+\ldots,\label{PMsetup2}
\end{gather}
where $\hat{H}$ is total Hamiltonian, $\hat{H}_0$ is unperturbed band Hamiltonian, $\mathbf{E}(\omega)=i\omega\mathbf{A}(\omega)$ is electric field and 
\begin{equation}
    h^{\mu}_{ab}= v^{\mu}_{ab} = \partial^{\mu}\epsilon_a \delta_{ab}+i\epsilon_{ab}A^{\mu}_{ab},
\end{equation} 
\begin{multline}
        h^{\mu\alpha}_{ab} = \partial^{\alpha}v^{\mu}_{ab}+iv^{\mu}_{ab}(A^{\alpha}_{bb}-A^{\alpha}_{aa})+\\+{{\sum_{c}}}^\prime\left\{ \frac{v^{\alpha}_{ac}v^{\mu}_{cb}}{\epsilon_{ca}}-\frac{v^{\mu}_{ac}v^{\alpha}_{cb}}{\epsilon_{bc}}\right\},
\end{multline}
\begin{multline}
        h^{\mu\alpha\beta}_{ab} = \partial^{\beta}h^{\mu\alpha}_{ab}+ih^{\mu\alpha}_{ab}(A^{\beta}_{bb}-A^{\beta}_{aa})+\\+{{\sum_{c}}}^\prime\left\{ \frac{v^{\beta}_{ac}h^{\mu\alpha}_{cb}}{\epsilon_{ca}}-\frac{h^{\mu\alpha}_{ac}v^{\beta}_{cb}}{\epsilon_{bc}}\right\},
\end{multline}    
where ${\sum}^\prime$ means that one omits terms when denominator vanishes and latin sub-indices, ($a,b,c$) denote band labels, while upper greek indices  ($\alpha, \beta, \gamma$) denote space components. 

From the above, the second order rectification conductivity is found to be:
\begin{multline}\label{park}
    \sigma^{\mu\alpha\beta}(0,\omega,-\omega)=\frac{e^3}{\omega^2}\sum_{abc}\Bigg\{f_a h^{\mu\alpha\beta}_{aa}+\\+f_{ab}\frac{h^{\alpha}_{ab}h^{\mu\beta}_{ba}}{\omega-\epsilon_{ab}}+f_{ab}\frac{h^{\beta}_{ab}h^{\mu\alpha}_{ba}}{-\omega-\epsilon_{ab}} +f_{ab}\frac{h^{\alpha\beta}_{ab}h^{\mu}_{ba}}{\epsilon_{ba}}+\\+\frac{h^{\alpha}_{ab}h^{\beta}_{bc}h^{\mu}_{ca}}{\epsilon_{ac}}\left(\frac{f_{ab}}{\omega-\epsilon_{ba}}-\frac{f_{cb}}{\omega-\epsilon_{bc}}\right)+\\+\frac{h^{\beta}_{ab}h^{\alpha}_{bc}h^{\mu}_{ca}}{\epsilon_{ac}}\left(\frac{f_{ab}}{-\omega-\epsilon_{ba}}-\frac{f_{cb}}{-\omega-\epsilon_{bc}}\right)\Bigg\}.
\end{multline}

Here $f_{ab}=f_a-f_b$, with $f_a$ the Fermi-Dirac occupation of band $a$. We will compare the expression above with the expression from Refs.\cite{PhysRevB.23.5590,PhysRevB.19.1548}. References \cite{PhysRevB.23.5590,PhysRevB.19.1548} employed the same gauge of Ref.\cite{PhysRevB.99.045121}, but assumed that the electrons have an underlying parabolic kinetic energy in addition to the periodic potential, and thus entirely neglects spin-orbit coupling effects. Therefore, the second order perturbed Hamiltonian for a single electron in Refs.\cite{PhysRevB.23.5590,PhysRevB.19.1548} has the form:
\begin{gather}
    \hat{H} = \hat{H}_0+\hat{V}(t),\qquad A^\alpha=\frac{E^\alpha_0}{\omega}\cos{\omega t}\label{KBsetup1}\\    
    \hat{V}(t) = -\frac{e}{m_0} \hat{p}^\alpha A^\alpha+\frac{e^2}{2m_0}A^\alpha A^\alpha\hat{\mathds{1}},\label{KBsetup3}
\end{gather}
where $m_0$ is the bare electron mass. The second order rectification conductivity of Refs.\cite{PhysRevB.23.5590,PhysRevB.19.1548} is therefore found to be:
\begin{multline}
    \sigma^{\mu}_{\alpha\beta}(\omega) = \frac{|e|^3}{8\pi^2\omega^2}\times\\\times\mathrm{Re}\left\{\sum_{\Omega=\pm\omega}\sum_{abc}f_{ba}\frac{v^{\alpha}_{ab}v^{\beta}_{bc}v^\mu_{ca}}{(\epsilon_{ac}-i\delta)(\epsilon_{ab}+\Omega-i\delta))}\right\}.
\end{multline}

For purposes of comparing both expressions it is convenient to split them into pieces. The conductivity from Refs.\cite{PhysRevB.23.5590,PhysRevB.19.1548} we split as follows:
\begin{gather}
    \mathrm{KB}_1 = \frac{e^3}{\omega^2}\mathrm{Re}\left[\sum_{abc}f_{ab}\frac{v^{\alpha}_{ab}v^{\beta}_{bc}v^\mu_{ca}}{(\epsilon_{ac}-i\delta)(\epsilon_{ab}+\omega-i\delta))}\right],\\
        \mathrm{KB}_2 = \frac{e^3}{\omega^2}\mathrm{Re}\left[\sum_{abc}f_{ab}\frac{v^{\alpha}_{ab}v^{\beta}_{bc}v^\mu_{ca}}{(\epsilon_{ac}-i\delta)(\epsilon_{ab}-\omega-i\delta))}\right],
\end{gather}
where KB stands for Kraut-Baltz from Refs.\cite{PhysRevB.23.5590,PhysRevB.19.1548}. We also split the conductivity from Eq.\eqref{park} (Ref.\cite{PhysRevB.99.045121}) into the following terms:
\begin{gather}
    \mathrm{PM}_0=\frac{e^3}{\omega^2}\sum_{ab}\Bigg\{f_a h^{\mu\alpha\beta}_{aa} +f_{ab}\frac{h^{\alpha\beta}_{ab}h^{\mu}_{ba}}{\epsilon_{ba}}\Bigg\},\\
    \mathrm{PM}_1 = \frac{e^3}{\omega^2}\sum_{ab}\Bigg\{ f_{ab}\frac{h^{\alpha}_{ab}h^{\mu\beta}_{ba}}{\omega-\epsilon_{ab}}+f_{ab}\frac{h^{\beta}_{ab}h^{\mu\alpha}_{ba}}{-\omega-\epsilon_{ab}}\Bigg\},\\
    \mathrm{PM}_2 =\frac{e^3}{\omega^2}\sum_{abc}\Bigg\{\frac{h^{\alpha}_{ab}h^{\beta}_{bc}h^{\mu}_{ca}}{\epsilon_{ac}}\left(\frac{f_{ab}}{\omega-\epsilon_{ba}}-\frac{f_{cb}}{\omega-\epsilon_{bc}}\right)\Bigg\},\\
    \mathrm{PM}_3 =\frac{e^3}{\omega^2}\sum_{abc}\Bigg\{\frac{h^{\beta}_{ab}h^{\alpha}_{bc}h^{\mu}_{ca}}{\epsilon_{ac}}\left(\frac{f_{ab}}{-\omega-\epsilon_{ba}}-\frac{f_{cb}}{-\omega-\epsilon_{bc}}\right)\Bigg\},
\end{gather}
where PM stands for Parker-Morimoto-Orenstein-Moore. Also we want to use one more index for each sub-term in these expressions. Namely, we will denote by $\mathrm{PM}_{1,1}$ the firs term from expression $\mathrm{PM}_{1}$, and analogously for other terms. 

By comparing Eqs.(\ref{PMsetup1},\ref{PMsetup2}) and Eqs.(\ref{KBsetup1}-\ref{KBsetup3}), we see that for systems considered in Refs.\cite{PhysRevB.23.5590,PhysRevB.19.1548} certain additional constrains are satisfied for the matrix elements as a consequence of the underlying parabolic dispersion. More specifically, the interband rectification conductivity from Refs.\cite{PhysRevB.23.5590,PhysRevB.19.1548}, can be obtained from the formula of Ref.\cite{PhysRevB.99.045121}, upon enforcing the following relations: 
\begin{gather}
    \hat{h}^\alpha = -\frac{\hbar}{m_0} \hat{p}^\alpha,\label{sumrules1}\\
    \hat{h}^{\alpha\beta} = \frac{\hbar^2}{e^2}\frac{e^2}{2m_0}\delta^{\alpha\beta}\hat{\mathds{1}},\\
    \hat{h}^{\alpha\beta\gamma}=\hat{h}^{\alpha\beta\gamma\delta}=\cdots=0.\label{sumrules2}
\end{gather}

From the expressions above one can see that the terms we have labeled PM$_0$ and PM$_1$ would vanish, because  the tensor $\hat{h}^{\alpha\beta}_{ab}\sim \delta_{ab}\delta^{\alpha\beta}$, is diagonal in the $(a,b)$ band indices, and therefore its contribution to the rectification conductivity vanishes after multiplying by the difference of Fermi occupation functions, $f_{ab}=f_a-f_b$, which are only non-zero for $a\neq b$. Now one can verify that the remaining terms agree between the two theories, specifically as follows:
\begin{multline}
    \mathrm{Re}\left[\mathrm{PM}_{21}\right] = \frac{e^3}{\omega^2}\mathrm{Re}\left[\sum_{abc}\frac{h^{\alpha}_{ab}h^{\beta}_{bc}h^{\mu}_{ca}}{\epsilon_{ac}}\frac{f_{ab}}{\omega-\epsilon_{ba}}\right] =\\=  \frac{e^3}{\omega^2}\mathrm{Re}\left[\sum_{abc}\frac{v^{\alpha}_{ab}v^{\beta}_{bc}v^{\mu}_{ca}}{\epsilon_{ac}}\frac{f_{ab}}{\epsilon_{ab}+\omega}\right] = \mathrm{KB}_1,
\end{multline}
\begin{multline}
    \mathrm{Re}\left[\mathrm{PM}_{32}\right]=-\frac{e^3}{\omega^2}\mathrm{Re}\left[\sum_{abc}\frac{h^{\beta}_{ab}h^{\alpha}_{bc}h^{\mu}_{ca}}{\epsilon_{ac}}\frac{f_{cb}}{-\omega-\epsilon_{bc}}\right]=\\=\left|\begin{array}{c}\text{complex}\\\text{conjugation}\end{array}\right|=-\frac{e^3}{\omega^2}\mathrm{Re}\left[\sum_{abc}\frac{h^{\beta}_{ba}h^{\alpha}_{cb}h^{\mu}_{ac}}{\epsilon_{ac}}\frac{f_{cb}}{-\omega-\epsilon_{bc}}\right]=\\= \left|a\leftrightarrow c\right|=-\frac{e^3}{\omega^2}\mathrm{Re}\left[\sum_{abc}\frac{h^{\beta}_{bc}h^{\alpha}_{ab}h^{\mu}_{ca}}{\epsilon_{ca}}\frac{f_{ab}}{-\omega-\epsilon_{ba}}\right]=\\=\frac{e^3}{\omega^2}\mathrm{Re}\left[\sum_{abc}\frac{h^{\beta}_{bc}h^{\alpha}_{ab}h^{\mu}_{ca}}{\epsilon_{ac}}\frac{f_{ab}}{\epsilon_{ab}-\omega}\right]=\\=\frac{e^3}{\omega^2}\mathrm{Re}\left[\sum_{abc}\frac{v^{\beta}_{bc}v^{\alpha}_{ab}v^{\mu}_{ca}}{\epsilon_{ac}}\frac{f_{ab}}{\epsilon_{ab}-\omega}\right]= \mathrm{KB}_2.
\end{multline}

Expressions for $\mathrm{Re}[$PM$]_{22}$ and $\mathrm{Re}[$PM$]_{31}$ can be obtained from $\mathrm{Re}[$PM$]_{21}$ and $\mathrm{Re}[$PM$]_{32}$ respectively by swapping indices $\alpha\leftrightarrow\beta$. Such a symmetrization automatically implied by conductivity once it is contracted with electric field indices, as is used for current expression in Ref.\cite{PhysRevB.23.5590}. Thus we conclude the conductivity from Ref.\cite{PhysRevB.23.5590} is equivalent to the sum of  terms $\mathrm{Re}[$PM$]_{2}$ and $\mathrm{Re}[$PM$]_{3}$. 

Note that in general (for example in the presence of spin-orbit coupling) $\mathrm{Re}[$PM$]_{0}$ and $\mathrm{Re}[$PM$]_{1}$ do contribute to the conductivity. These terms include not only intra-band terms, but also some contribution to interband shift currents terms that are missing in Refs.\cite{PhysRevB.23.5590,PhysRevB.19.1548}, and are given by:    
\begin{multline}
    \mathrm{Re}\left[\frac{e^3}{\omega^2}\sum_{ab} f_{ab}\frac{iv^{\alpha}_{ab}v^{\mu}_{ba}(A^{\beta}_{aa}-A^{\beta}_{bb})}{\omega-\epsilon_{ab}}\right]+\\+\mathrm{Re}\left[\frac{e^3}{\omega^2}\sum_{ab} f_{ab}\frac{v^{\alpha}_{ab}}{\omega-\epsilon_{ab}}{{\sum_{c}}}^\prime\left\{ \frac{v^{\beta}_{bc}v^{\mu}_{ca}}{\epsilon_{cb}}-\frac{v^{\mu}_{bc}v^{\beta}_{ca}}{\epsilon_{ac}}\right\}\right]+\\+\left(\begin{array}{c}\alpha\leftrightarrow\beta  \\\omega\leftrightarrow-\omega\end{array}\right)
\end{multline}

Or in notations used in Refs.\cite{PhysRevB.23.5590,PhysRevB.19.1548}:
\begin{multline}
    \mathrm{Re}\left[\frac{e^3}{2\omega^2}\sum_{\Omega=\pm\omega}\sum_{ab} f_{ab}\frac{iv^{\alpha}_{ab}v^{\mu}_{ba}(A^{\beta}_{aa}-A^{\beta}_{bb})}{\Omega+\epsilon_{ba}-i\delta}\right]+\\+\mathrm{Re}\left[\frac{e^3}{\omega^2}\sum_{\Omega=\pm\omega}\sum_{ab}{{\sum_{c}}}^\prime f_{ab}\frac{v^{\alpha}_{ab}}{\Omega+\epsilon_{ba}-i\delta} \frac{v^{\beta}_{bc}v^{\mu}_{ca}}{\epsilon_{cb}}\right]+\\+\left(\begin{array}{c}\alpha\leftrightarrow\beta  \end{array}\right)
\end{multline}

To show that the extra shift current contribution, which we demonstrate above, is crucial - we will compute our Quantum Rectification Sum Rule for shift currents described in Refs.\cite{PhysRevB.23.5590,PhysRevB.19.1548} and Ref.\cite{PhysRevB.99.045121}. With the proper regularization of the infinitesimal imaginary parts of the frequency denominators of both of these references, the frequency integration of interband conductivities are:
\begin{multline}\label{PMSR}
    2\frac{\hbar^2}{e^3}\frac{1}{\pi}\int_0^{\infty} \mathrm{Re}\left[\mathrm{PM}^{\mu\beta \alpha}(0,-\omega,\omega)\right] d\omega =\\= \sum_af_{a}\left(i[A^{\beta}\partial^\alpha A^{\mu}]_{aa}+[A^{\beta} ,[A^\alpha,\bar{A}^{\mu}]_{aa}\right)+\\+\left(\begin{array}{c}\alpha\leftrightarrow\beta  \end{array}\right),
\end{multline}
\begin{multline}\label{YSSR}
    2\frac{\hbar^2}{e^3}\frac{1}{\pi}\int_0^{\infty} \mathrm{Re}\left[\mathrm{KB}^{\gamma\beta \alpha}(0,-\omega,\omega)\right]d\omega =\\= \sum_{abc}f_{ab}\frac{\epsilon_{bc}}{\epsilon_{ab}}\mathrm{Re}\left[A^{\alpha}_{ab}A^{\beta}_{bc}A^{\mu}_{ca}\right].
\end{multline}

Therefore we see that while the expression for interband contribution to the sum rule obtained from Ref.\cite{PhysRevB.99.045121} (Eq.(\ref{PMSR})) is the same as the one derived in our previous paper Ref.\cite{PhysRevLett.123.246602}, the one that would be derived from Refs.\cite{PhysRevB.23.5590,PhysRevB.19.1548} (Eq.(\ref{YSSR})) differs from them. As previously discussed, this is because  Refs.\cite{PhysRevB.23.5590,PhysRevB.19.1548} are missing some terms which only vanish if one assumes a parabolic dispersion which completely neglects the spin-orbit coupling.

\bibliography{mysuperbib}

\end{document}